\documentclass[12pt]{article}
\usepackage{a4}
\usepackage{amsmath}
\usepackage{amssymb}
\usepackage{amsfonts}
\usepackage{graphics}
\usepackage{epsfig}
\usepackage{latexsym}
\usepackage{epic}
\usepackage{eepic}

\begin{document}

\noindent hep-th/
   \hfill  September 2010 \\

\renewcommand{\theequation}{\arabic{section}.\arabic{equation}}
\thispagestyle{empty}
\vspace*{-1,5cm}
\noindent \vskip3.3cm

\begin{center}
{\Large\bf A generating function for the cubic interactions of higher spin fields}

{\large Ruben Manvelyan ${}^{\dag\ddag}$, Karapet Mkrtchyan${}^{\dag\ddag}$ \\and Werner R\"uhl
${}^{\dag}$}
\medskip

${}^{\dag}${\small\it Department of Physics\\ Erwin Schr\"odinger Stra\ss e \\
Technical University of Kaiserslautern, Postfach 3049}\\
{\small\it 67653
Kaiserslautern, Germany}\\
\medskip
${}^{\ddag}${\small\it Yerevan Physics Institute\\ Alikhanian Br.
Str.
2, 0036 Yerevan, Armenia}\\
\medskip
{\small\tt manvel,ruehl@physik.uni-kl.de; karapet@yerphi.am}
\end{center}\vspace{2cm}

\bigskip
\begin{center}
{\sc Abstract}
\end{center}
\quad
We present an off-shell generating function for all cubic interactions of Higher Spin gauge fields constructed in \cite{GTI}. It is a generalization of the on-shell generating function proposed in \cite{ST}, is written in a very compact way, and turns out to have a remarkable structure.

\newpage

\section{Introduction and notations}

\quad Higher Spin gauge field theory is one of the most important and puzzling problems in modern quantum field theory.
It is a subject of many articles and always stays in the center of attention during last thirty years. Despite the fact that consistent equations of motion for Higher Spin gauge fields are known over twenty years \cite{VasilievEqn}, the question of existence of Lagrangians for interacting Higher Spin gauge fields is still open. The subject of special interest is a minimal selfinteraction of even spin gauge fields, where one can naively expect the existence of an Einstein-Hilbert type nonlinear action for any single even spin gauge field. Although there are known restrictions on Higher Spin theories in flat space-time, the recent development \cite{GTI} has shown that \emph{there is a local higher derivative cubic Lagrangian for gauge fields with any higher spins}. This shifts the no-go theorems to the quartic power of fields in interaction Lagrangians, where one can expect the final battle for the existence of local (or nonlocal) Lagrangians for interacting HS gauge field theory in flat space.

\quad  The free Lagrangian for Higher Spin gauge fields both in flat space and in constantly curved backgrounds (dS and AdS) are known over thirty years \cite{Frons}. In contrast to free theory, attempts to construct Lagrangians for interacting theories haven't been successful yet beyond the cubic vertices. In this letter we are going to discuss only trilinear interactions of Higher Spin gauge fields.

\quad Our recent results \cite{GTI}, \cite{MMR0}-\cite{MR} on Higher Spin gauge field cubic interactions in flat space, which certainly reproduce the flat limit of the Fradkin-Vasiliev vertex for higher spin coupling to gravity \cite{Vasiliev}, show that all interactions of higher spin gauge fields with any spins $s_{1}, s_{2}, s_{3}$ both in flat space and in dS or AdS are unique\footnote{The cubic interaction Lagrangian is unique up to partial integration and field redefinition.}. This was already proven for some low spin cases of both the Fradkin-Vasiliev vertex for $2, s, s$ and the nonabelian vertex for $1, s, s$  in \cite{boulanger}.

\quad The first important step towards cubic interactions in Higher Spin gauge field theory was done in 1984 by Berends, Burgers and van Dam \cite{vanDam}. They constructed a cubic selfinteraction Lagrangian for spin three gauge fields and proved impossibility of extension to higher orders. Their arguments are based on gauge algebra, which does not close for a single spin three nonabelian field. The authors give an optimistic hope that it will be possible to extend this Lagrangian to higher orders if one takes into account corrections from interactions with gauge fields with spins higher than three. A recent discussion on this subject appeared from Bekaert, Boulanger and Leclerq \cite{bek}. They show the impossibility to close this (spin 3) algebra taking into account corrections from interactions of other fields with spins higher (or lower) than three. It is not yet known whether there is or there is not any nonlinear selfinteraction for a spin three gauge field in the background space with nonzero cosmological constant (pure spin three theory).

\quad Another important step was done by Fradkin and Vasiliev in \cite{Vasiliev} where coupling of Higher Spin gauge fields to linearized gravity was constructed in the constantly curved background. The interesting property of this Lagrangian is it's non-analyticity in the cosmological constant, therefore excluding a flat space limit. However it was shown already  in \cite{boulanger} that after rescaling of Higher Spin gauge fields one can observe a flat limit for the Fradkin-Vasiliev interactions. In our approach the spin $s$ gauge field has scaling dimension $[length]^{s-2}$, therefore the Fradkin-Vasiliev vertex has a flat limit with $2s-2$ derivatives (minimal possible number) in the $2-s-s$ interaction which has the same scaling dimension as the Einstein-Hilbert Lagrangian terms. As it was shown by Metsaev in \cite{Metsaev} using a light cone gauge approach, there are three different couplings to linearized gravity with different numbers of derivatives for any higher spin $s$ field, and in general $min\{s_{1},s_{2},s_{3}\}+1$ different possibilities with different numbers of derivatives for the $s_{1}-s_{2}-s_{3}$ interaction. All these interactions were derived in a covariant off-shell formulation in \cite{GTI}.

\quad In the recent paper \cite{ST} by Sagnotti and Taronna the authors proposed an on-shell generating function for the general interaction presented in \cite{GTI} from a string theory consideration. In this article we are going to present an off-shell extension of that generating function which can surprisingly be enhanced with a beautiful Grassmann structure, the string origin of which is not clear yet.

\quad For some important results on higher spin cubic interactions see \cite{Vasiliev1}-\cite{Zinoviev} and references therein. For recent reviews see \cite{review}. We should note also that a connection between non-Abelian tensor gauge theories and string amplitudes was also explored in  \cite{Savvidy}.

\quad In this paper we consider a Higher Spin gauge field theory in Fronsdal's formulation. The spin $s$ field is a rank $s$ symmetric, double traceless tensor and we consider here only one copy for any spin, so these interactions are for even spin gauge field (self)interactions only.

\quad To continue with this subject we introduce here briefly our standard notations coming from our previous papers about HSF (see for example \cite{MMR1}). As usual we utilize instead of symmetric tensors such as $h^{(s)}_{\mu_1\mu_2...\mu_s}(z)$ the homogeneous polynomials in the vector $a^{\mu}$ of degree $s$ at the base point $z$
\begin{equation}
h^{(s)}(z;a) = \sum_{\mu_{i}}(\prod_{i=1}^{s} a^{\mu_{i}})h^{(s)}_{\mu_1\mu_2...\mu_s}(z) .
\end{equation}
Then we can write the symmetrized gradient, trace and divergence \footnote{To distinguish easily between "a" and "z" spaces we introduce the notation $\nabla_{\mu}$ for space-time derivatives $\frac{\partial}{\partial z^{\mu}}$.}
\begin{eqnarray}
&&Grad:h^{(s)}(z;a)\Rightarrow Gradh^{(s+1)}(z;a) = (a\nabla)h^{(s)}(z;a) , \\
&&Tr:h^{(s)}(z;a)\Rightarrow Trh^{(s-2)}(z;a) = \frac{1}{s(s-1)}\Box_{a}h^{(s)}(z;a) ,\\
&&Div:h^{(s)}(z;a)\Rightarrow Divh^{(s-1)}(z;a) = \frac{1}{s}(\nabla\partial_{a})h^{(s)}(z;a) .
\end{eqnarray}

\quad Here we only present
Fronsdal's Lagrangian in terms of these conventions\footnote{From now on we will presuppose integration everywhere where it is necessary (we work with a Lagrangian as with an action) and therefore we will neglect all $d$ dimensional space-time total derivatives when making a partial integration.}:
\begin{equation}\label{1.42(1)}
 \mathcal{L}_{0}(h^{(s)}(a))=-\frac{1}{2}h^{(s)}(a)*_{a}\mathcal{F}^{(s)}(a)
    +\frac{1}{8s(s-1)}\Box_{a}h^{(s)}(a)*_{a}\Box_{a}\mathcal{F}^{(s)}(a) ,
\end{equation}
where $\mathcal{F}^{(s)}(z;a)$ is the Fronsdal tensor
\begin{eqnarray}
\mathcal{F}^{(s)}(z;a)=\Box h^{(s)}(z;a)-s(a\nabla)D^{(s-1)}(z;a) , \quad\label{0.32(1)}
\end{eqnarray}
and $D^{(s-1)}(z;a)$ is the deDonder tensor or traceless divergence of the higher spin gauge field
\begin{eqnarray}\label{0.42(D)}
 && D^{(s-1)}(z;a) = Divh^{(s-1)}(z;a)
-\frac{s-1}{2}(a\nabla)Trh^{(s-2)}(z;a) ,\\
&& \Box_{a} D^{(s-1)}(z;a)=0 .
\end{eqnarray}
The initial gauge variation of order zero in the spin $s$ field is
\begin{eqnarray}\label{0.5}
\delta_{(0)} h^{(s)}(z;a)=s (a\nabla)\epsilon^{(s-1)}(z;a) ,
\end{eqnarray}
with a traceless gauge parameter for the double traceless gauge field
\begin{eqnarray}
&&\Box_{a}\epsilon^{(s-1)}(z;a)=0 ,\label{0.6}\\
&&\Box_{a}^{2}h^{(s)}(z;a)=0 .
\end{eqnarray}

\section{Free Lagrangian for all higher spin gauge fields}
\setcounter{equation}{0}
\quad We introduce a generating function for HS gauge fields by
\begin {equation}
\Phi(z;a) = \sum_{s=0}^{\infty} \frac{1}{s!}h^{(s)}(z;a) \label{sf}
\end{equation}
where we assume that the spin $s$ field has scaling dimension $s-2$, the $a_{i}$ vectors have dimension $-1$, and therefore all terms in the generating function for higher spin gauge fields (\ref{sf}) have the same dimension $-2$.

A zeroth order gauge transformation for this field reads as
\begin{eqnarray}
\delta^{0}_{\Lambda}\Phi(z;a) = (a\nabla)\Lambda(z;a)\label{gv},\\
\delta^{0}_{\Lambda}D_{a}\Phi(z;a) = \Box \Lambda(z;a),\\
\delta^{0}_{\Lambda}\Box_{a}\Phi(z;a) = 2(\nabla\partial_{a})\Lambda(z;a).
\end{eqnarray}
where
\begin{eqnarray}
\Lambda(z;a)=\sum_{s=1}^{\infty} \frac{1}{(s-1)!}\epsilon^{(s-1)}(z;a)\label{gp},
\end{eqnarray}
is the generating function of the gauge parameters and is dimensionless\footnote{The gauge parameter for spin $s$ field $\epsilon^{(s-1)}$ has scaling dimension $s-1$, therefore after contraction with $s-1$ $a$-vectors becomes dimensionless.}.

Fronsdal's constraint for the gauge parameter reads as
\begin{eqnarray}
\Box_{a}\Lambda(z;a)=0\label{FC},
\end{eqnarray}
For a spin $s$ field gauge variation we get as expected
\begin{eqnarray}
\delta_{\epsilon}^{0}h^{(s)}(z;a)=s(a\nabla)\epsilon^{(s-1)}(z;a)\label{gvs},
\end{eqnarray}

The second Fronsdal constraint of the gauge field reads in these notations
\begin{eqnarray}
\Box_{a}^{2}\Phi(z;a)=0\label{FC2},
\end{eqnarray}

We introduced the "de Donder" operator
\begin{equation}
D_{a_i} = (\partial_{a_i}\nabla_i)-\frac{1}{2}(a_i\nabla_i)\Box_{a_i} \label{D}
\end{equation}
This operator is "linear" in $\partial_{a_i}$.

Here we write the quadratic Lagrangian for free higher spin gauge fields in general form using the generating function for HS fields (\ref{sf}).
First we introduce Fronsdal's operator
\begin{eqnarray}
\mathcal{F}_{a_{i}}=\Box_{i}-(a_{i}\nabla_{i})(\nabla_{i}\partial_{a_{i}})+\frac{1}{2}(a_{i}\nabla_{i})^{2}\Box_{a_{i}},
\end{eqnarray}
or with the help of (\ref{D})
\begin{eqnarray}
\mathcal{F}_{a_{i}}=\Box_{i}-(a_{i}\nabla_{i})D_{a_{i}}.
\end{eqnarray}
The operator of the equation of motion can be written in the form
\begin{eqnarray}
\mathcal{G}_{a_{i}}=\mathcal{F}_{a_{i}}-\frac{a_{i}^{2}}{4}\Box_{a_{i}}\mathcal{F}_{a_{i}}
\end{eqnarray}
Now we can write the free Lagrangian for all gauge fields of any spin in a symmetric elegant form
\begin{eqnarray}
\mathcal{L}^{\emph{free}}(z)&=&\frac{\kappa}{2} exp[\lambda^{2}\partial_{a_{1}}\partial_{a_{2}}]\int_{z_{1}z_{2}} \delta (z_{1}-z) \delta(z_{2}-z)\nonumber\\
                            &&\{(\nabla_{1}\nabla_{2})-\lambda^{2}D_{a_1}D_{a_2}-\frac{\lambda^{4}}{4}(\nabla_{1}\nabla_{2})\Box_{a_1}\Box_{a_2}\}
                            \Phi(z_{1};a_{1})\Phi(z_{2};a_{2})\mid _{a_{1}=a_{2}=0}\ \ \ \
\end{eqnarray}
where $\lambda$ has scaling dimension $-1$, therefore $\lambda^{2}$ compensates the dimension of the operator in the exponent.
We will see that all relative coupling constants of HS interactions can be expressed as powers of $\lambda$. The parameter $\kappa$ is a constant which makes the action dimensionless (analogous to Einstein's constant and simply connected with the latter). It has scaling dimension $6-d$, where $d$ is the space-time dimension. For Einstein's constant $\kappa_{E}$ we get
\begin{equation}
\kappa_{E}^{-2}=\kappa \lambda^{4}
\end{equation}
It is now obvious that in the free Lagrangian there is no mixing between gauge fields of different spin.
It can also be written in such forms
\begin{eqnarray}
\mathcal{L}^{\emph{free}}(z)&=&-\frac{1}{2}exp[\lambda^{2}\partial_{a_{1}}\partial_{a_{2}}]\int_{z_{1}} \delta(z_{1}-z)
                                (\mathcal{G}_{a_{1}})\Phi(z_{1};a_{1})\Phi(z;a_{2})\mid _{a_{1}=a_{2}=0}\nonumber\\
                            &=&-\frac{1}{2}exp[\lambda^{2}\partial_{a_{1}}\partial_{a_{2}}]\int_{z_{2}} \delta(z_{2}-z)
                                (\mathcal{G}_{a_{2}})\Phi(z;a_{1})\Phi(z_{2};a_{2})\mid _{a_{1}=a_{2}=0}
\end{eqnarray}
These expressions reproduce Fronsdal's Lagrangians for all gauge fields with any spin.

\section{Cubic Interactions}
\setcounter{equation}{0}
\quad We are going to present a very beautiful and compact form of all HS gauge field interactions derived in \cite{GTI}.
First we rewrite the leading term of a general trilinear interaction of higher spin gauge fields with any spins $s_{1},s_{2},s_{3}$
\footnote{$\nabla_{2}\partial_{a}=\frac{\partial}{\partial a^{\mu}}\nabla_{2}^{\mu}$ and so on.}
\begin{eqnarray}
&&\mathcal{L}^{leading}_{(1)}(h^{(s_{1})}(z),h^{(s_{2})}(z),h^{(s_{3})}(z))\nonumber\\
&&=\sum_{\alpha+\beta+\gamma = n}\frac{1}{\alpha!\beta!\gamma!}\int_{z_{1},z_{2},z_{3}} \delta(z-z_{1})\delta(z-z_{2})\delta(z-z_{3})\nonumber\\
&&\left[(\nabla_{1}\partial_{c})^{s_{3}-n+\gamma}(\nabla_{2}\partial_{a})^{s_{1}-n+\alpha}(\nabla_{3}\partial_{b})^{s_{2}-n+\beta}
(\partial_{a}\partial_{b})^{\gamma}(\partial_{b}\partial_{c})^{\alpha}(\partial_{c}\partial_{a})^{\beta}\right]\nonumber\\
&&h^{(s_{1})}(a;z_{1})h^{(s_{2})}(b;z_{2})h^{(s_{3})}(c;z_{3}) ,\label{leading}
\end{eqnarray}
where the number of derivatives is
\begin{eqnarray}
\Delta=s_{1}+s_{2}+s_{3}-2n,\\
0\leq n\leq min(s_{1},s_{2},s_{3})
\end{eqnarray}
As we see, the minimal and maximal possible numbers of derivatives are
\begin{eqnarray}
\Delta_{min}&=&s_{1}+s_{2}+s_{3}-2min(s_{1},s_{2},s_{3}),\\
\Delta_{max}&=&s_{1}+s_{2}+s_{3}.
\end{eqnarray}
These nteractions trivialize only if we have two equal spin values and the third value is odd.
This we call the $\ell-s-s$ case, where $\ell$ is odd.
In that case we should have a multiplet of spin $s$ fields, with at least two charges to couple to the spin $\ell$ field.
In the case of $\ell-\ell-\ell$ odd spin self interaction, the number of possible charges in the multiplet should be at least 3.
The case of $\Delta_{min}$ is important also because only in that case the interaction (\ref{leading}) has the same dimension as the lowest spin field free Lagrangian.

The same Lagrangian can be written in the following way (due to a constant normalization factor $2^{\Delta}$)
\begin{eqnarray}
&&\mathcal{L}^{leading}_{(1)}(h^{(s_{1})}(z),h^{(s_{2})}(z),h^{(s_{3})}(z))\nonumber\\
&&=\sum_{\alpha+\beta+\gamma = n}\frac{1}{\alpha!\beta!\gamma!}\int_{z_{1},z_{2},z_{3}} \delta(z-z_{1})\delta(z-z_{2})\delta(z-z_{3})\nonumber\\
&&\left[(\nabla_{12}\partial_{c})^{s_{3}-n+\gamma}(\nabla_{23}\partial_{a})^{s_{1}-n+\alpha}(\nabla_{31}\partial_{b})^{s_{2}-n+\beta}
(\partial_{a}\partial_{b})^{\gamma}(\partial_{b}\partial_{c})^{\alpha}(\partial_{c}\partial_{a})^{\beta}\right]\nonumber\\
&&h^{(s_{1})}(a;z_{1})h^{(s_{2})}(b;z_{2})h^{(s_{3})}(c;z_{3}) ,\label{interaction}
\end{eqnarray}
where
\begin{eqnarray}
&&\nabla_{12}=\nabla_{1}-\nabla_{2},\\
&&\nabla_{23}=\nabla_{2}-\nabla_{3},\\
&&\nabla_{31}=\nabla_{3}-\nabla_{1}.
\end{eqnarray}
Now we can see that the following expression is a generating function for the leading term of all interactions of HS gauge fields.
\begin{eqnarray}
&\mathcal{A}^{00} = \int_{z_{1},z_{2},z_{3}} \delta(z-z_{1})\delta(z-z_{2})\delta(z-z_{3})exp W \nonumber \\
&\Phi_1(z_1;a_{1}+\frac{1}{2}\nabla_{23})\Phi_2(z_2;a_{2}+\frac{1}{2}\nabla_{31})\Phi_3(z_3;a_{3}+\frac{1}{2}\nabla_{12})\mid_{a_{1}=a_{2}=a_{3}=0}\label{GenF}
\end{eqnarray}
with
\begin{eqnarray}
W = \frac{\lambda^{2}}{2}[(\partial_{a_{1}}\partial_{a_{2}})(\partial_{a_{3}}\nabla_{12})+(\partial_{a_{2}}\partial_{a_{3}})(\partial_{a_{1}}\nabla_{23})
+(\partial_{a_{3}}\partial_{a_{1}})(\partial_{a_{2}}\nabla_{31})]\qquad \label{W1}
\end{eqnarray}
This can be written in another form
\begin{eqnarray}
\mathcal {A}^{00}(\Phi(z)) = \int_{z_{1},z_{2},z_{3}} \delta(z-z_{1,2,3}) exp \hat W \times\Phi(z_1;a_{1})\Phi(z_2;a_{2})\Phi(z_3;a_{3})\mid_{a_{1}=a_{2}=a_{3}=0}\nonumber\\\label{exp}
\end{eqnarray}
where
\begin{eqnarray}
&\hat W = \frac{\lambda^{2}}{2}[(\partial_{a_{1}}\partial_{a_{2}})(\partial_{a_{3}}\nabla_{12})+(\partial_{a_{2}}\partial_{a_{3}})(\partial_{a_{1}}\nabla_{23})
+(\partial_{a_{3}}\partial_{a_{1}})(\partial_{a_{2}}\nabla_{31})]\nonumber \\
&+\frac{1}{2}[(\partial_{a_{3}}\nabla_{12})+(\partial_{a_{1}}\nabla_{23})+(\partial_{a_{2}}\nabla_{31})]\label{W},\\
&\int_{z_{1},z_{2},z_{3}} \delta(z-z_{1,2,3})=\int_{z_{1},z_{2},z_{3}} \delta(z-z_{1})\delta(z-z_{2})\delta(z-z_{3})
\end{eqnarray}
for brevity. Furthermore we will always assume this integration with delta functions, without writing it explicitly.
The operator in the second row of (\ref{W}) is a dimensionless operator, therefore it does not need any dimensional constant multiplier.

Now we can derive all other terms in the Lagrangian using the following important relation
\begin{eqnarray}
[exp\hat W, A]=exp\hat W [\hat W, A]+exp\hat W [\hat W, [\hat W, A]]+exp\hat W [\hat W, [\hat W, [\hat W, A]]]+...
\end{eqnarray}
for any operator $A$. And therefore
\begin{eqnarray}
&&[exp\hat W, (a_{1}\nabla_{1})]=exp\hat W [\hat W, (a_{1}\nabla_{1})],\\
&&[exp\hat W, (a_{2}\nabla_{2})]=exp\hat W [\hat W, (a_{2}\nabla_{2})],\\
&&[exp\hat W, (a_{3}\nabla_{3})]=exp\hat W [\hat W, (a_{3}\nabla_{3})].
\end{eqnarray}
The following commutators will be used many times while deriving trace and divergence terms
\begin{eqnarray}
&&[\hat W, (a_{1}\nabla_{1})]=-\frac{\lambda^{2}}{4}[(\partial_{a_{2}}\nabla_{2})(\partial_{a_{3}}\nabla_{12})
                                                    +(\partial_{a_{3}}\nabla_{3})(\partial_{a_{2}}\nabla_{31})]
                              +\frac{1}{2}[\lambda^{2}(\partial_{a_{2}}\partial_{a_{3}})+1]\nabla_{1}\nabla_{23},\qquad\\
&&[\hat W, (a_{2}\nabla_{2})]=-\frac{\lambda^{2}}{4}[(\partial_{a_{3}}\nabla_{3})(\partial_{a_{1}}\nabla_{23})
                                                    +(\partial_{a_{1}}\nabla_{1})(\partial_{a_{3}}\nabla_{12})]
                              +\frac{1}{2}[\lambda^{2}(\partial_{a_{3}}\partial_{a_{1}})+1]\nabla_{2}\nabla_{31},\qquad\\
&&[\hat W, (a_{3}\nabla_{3})]=-\frac{\lambda^{2}}{4}[(\partial_{a_{1}}\nabla_{1})(\partial_{a_{2}}\nabla_{31})
                                                    +(\partial_{a_{2}}\nabla_{2})(\partial_{a_{1}}\nabla_{23})]
                              +\frac{1}{2}[\lambda^{2}(\partial_{a_{1}}\partial_{a_{2}})+1]\nabla_{3}\nabla_{12}.\qquad
\end{eqnarray}
Note that
\begin{eqnarray}
&&\nabla_{1}\nabla_{23}=\Box_{3}-\Box_{2},\\
&&\nabla_{2}\nabla_{31}=\Box_{1}-\Box_{3},\\
&&\nabla_{3}\nabla_{12}=\Box_{2}-\Box_{1},
\end{eqnarray}
which is obvious because\footnote{We always understand partial integrations to be performed, working with a Lagrangian as with an action.}
\begin{eqnarray}
\nabla_{1}+\nabla_{2}+\nabla_{3}=0.
\end{eqnarray}
We are working with the same type of diagram as in \cite{GTI}.
\begin{eqnarray}
\setlength{\unitlength}{0.254mm}
\begin{picture}(280,275)(125,-355)
        \allinethickness{0.254mm}\path(405,-80)(405,-175) 
        \allinethickness{0.254mm}\path(345,-235)(345,-80) 
        \allinethickness{0.254mm}\path(285,-295)(285,-80) 
        \allinethickness{0.254mm}\path(225,-355)(225,-80) 
        \allinethickness{0.254mm}\path(125,-80)(125,-355) 
        \allinethickness{0.254mm}\path(165,-80)(165,-355) 
        \allinethickness{0.254mm}\path(405,-80)(125,-80) 
        \allinethickness{0.254mm}\path(405,-175)(125,-175) 
        \allinethickness{0.254mm}\path(345,-235)(125,-235) 
        \allinethickness{0.254mm}\path(285,-295)(125,-295) 
        \allinethickness{0.254mm}\path(225,-355)(125,-355) 
        \allinethickness{0.254mm}\path(125,-115)(405,-115) 
        \allinethickness{0.254mm}\path(125,-80)(165,-115) 
        \put(130,-110){\shortstack{$\Box_{a_{i}}$}} 
        \put(141,-94){\shortstack{$D_{a_{i}}$}} 
        \put(190,-106){\shortstack{$0$}} 
        \put(250,-106){\shortstack{$1$}} 
        \put(310,-106){\shortstack{$2$}} 
        \put(370,-106){\shortstack{$3$}} 
        \put(140,-151){\shortstack{$0$}} 
        \put(140,-211){\shortstack{$1$}} 
        \put(140,-271){\shortstack{$2$}} 
        \put(140,-331){\shortstack{$3$}} 
        \put(185,-151){\shortstack{$\mathcal {A}^{00}$}} 
        \put(235,-151){\shortstack{$\mathcal {A}^{10}$}} 
        \put(295,-151){\shortstack{$\mathcal {A}^{20}$}} 
        \put(355,-151){\shortstack{$\mathcal {A}^{30}$}} 
        \put(185,-211){\shortstack{$\mathcal {A}^{01}$}} 
        \put(185,-271){\shortstack{$\mathcal {A}^{02}$}} 
        \put(185,-331){\shortstack{$\mathcal {A}^{03}$}} 
        \put(235,-211){\shortstack{$\mathcal {A}^{11}$}} 
        \put(295,-211){\shortstack{$\mathcal {A}^{21}$}} 
        \put(235,-271){\shortstack{$\mathcal {A}^{12}$}} 
\end{picture}
\end{eqnarray}
Now we take a gauge variation of $\mathcal {A}^{00}$, and find generating functions for all other terms in the cubic Lagrangian.  A simple but elegant structure is exhibited by the first row of the diagram
\begin{eqnarray}
\mathcal {A}^{10}(\Phi(z))=\mathcal {A}^{30}(\Phi(&z&))=0,\\
\mathcal {A}^{20}(\Phi(z)) = \frac{1}{4} exp\hat W \{&&+[\lambda^{2}(\partial_{a_{1}}\partial_{a_{2}})+1][\lambda^{2}(\partial_{a_{2}}\partial_{a_{3}})+1]D_{a_{3}}D_{a_{1}}\nonumber\\
   &&+[\lambda^{2}(\partial_{a_{2}}\partial_{a_{3}})+1][\lambda^{2}(\partial_{a_{3}}\partial_{a_{1}})+1]D_{a_{1}}D_{a_{2}}\nonumber\\
   &&+[\lambda^{2}(\partial_{a_{3}}\partial_{a_{1}})+1][\lambda^{2}(\partial_{a_{1}}\partial_{a_{2}})+1]D_{a_{2}}D_{a_{3}}\}\nonumber\\
                                      &&\Phi(z_1;a_{1})\Phi(z_2;a_{2})\Phi(z_3;a_{3})\mid_{a_{1}=a_{2}=a_{3}=0}\qquad\qquad
\end{eqnarray}

Other terms are

\begin{eqnarray}
\mathcal {A}^{01}(\Phi(z))=0,&&\\
\mathcal {A}^{11}(\Phi(z)) = \frac{\lambda^{2}}{16}exp\hat W
\{&&+[\lambda^{2}(\partial_{a_{1}}\partial_{a_{2}})+1](\partial_{a_{1}}\nabla_{23})\Box_{a_{3}}D_{a_{2}}\nonumber\\
&&-[\lambda^{2}(\partial_{a_{1}}\partial_{a_{2}})+1](\partial_{a_{2}}\nabla_{31})\Box_{a_{3}}D_{a_{1}}\nonumber\\
&&+[\lambda^{2}(\partial_{a_{2}}\partial_{a_{3}})+1](\partial_{a_{2}}\nabla_{31})\Box_{a_{1}}D_{a_{3}}\nonumber\\
&&-[\lambda^{2}(\partial_{a_{2}}\partial_{a_{3}})+1](\partial_{a_{3}}\nabla_{12})\Box_{a_{1}}D_{a_{2}}\nonumber\\
&&+[\lambda^{2}(\partial_{a_{3}}\partial_{a_{1}})+1](\partial_{a_{3}}\nabla_{12})\Box_{a_{2}}D_{a_{1}}\nonumber\\
&&-[\lambda^{2}(\partial_{a_{3}}\partial_{a_{1}})+1](\partial_{a_{1}}\nabla_{23})\Box_{a_{2}}D_{a_{3}}\}\nonumber\\
&&\Phi(z_1;a_{1})\Phi(z_2;a_{2})\Phi(z_3;a_{3})\mid_{a_{1}=a_{2}=a_{3}=0}
\end{eqnarray}
and so on.

All these expressions can be written in a very elegant form. First we introduce Grassmann variables by

\begin{eqnarray}
\eta_{a_{1}}, \bar{\eta}_{a_{1}}, \eta_{a_{2}}, \bar{\eta}_{a_{2}}, \eta_{a_{3}}, \bar{\eta}_{a_{3}}.
\end{eqnarray}

Then we change expressions in the formula (\ref{exp}) in a following way

\begin{eqnarray}
&&(\partial_{a_{1}}\partial_{a_{2}}) \rightarrow (\partial_{a_{1}}\partial_{a_{2}})+\frac{1}{4}\eta_{a_{1}}\bar{\eta}_{a_{2}}\Box_{a_{2}}+\frac{1}{4}\eta_{a_{2}}\bar{\eta}_{a_{1}}\Box_{a_{1}},\\
&&(\partial_{a_{2}}\partial_{a_{3}}) \rightarrow (\partial_{a_{2}}\partial_{a_{3}})+\frac{1}{4}\eta_{a_{2}}\bar{\eta}_{a_{3}}\Box_{a_{3}}+\frac{1}{4}\eta_{a_{3}}\bar{\eta}_{a_{2}}\Box_{a_{2}},\\
&&(\partial_{a_{3}}\partial_{a_{1}}) \rightarrow (\partial_{a_{3}}\partial_{a_{1}})+\frac{1}{4}\eta_{a_{3}}\bar{\eta}_{a_{1}}\Box_{a_{1}}+\frac{1}{4}\eta_{a_{1}}\bar{\eta}_{a_{3}}\Box_{a_{3}},\\
&&(\partial_{a_{1}}\nabla_{23}) \rightarrow (\partial_{a_{1}}\nabla_{23})+\eta_{a_{1}}\bar{\eta}_{a_{2}}D_{a_{2}}-\eta_{a_{1}}\bar{\eta}_{a_{3}}D_{a_{3}}\\
&&(\partial_{a_{2}}\nabla_{31}) \rightarrow (\partial_{a_{2}}\nabla_{31})+\eta_{a_{2}}\bar{\eta}_{a_{3}}D_{a_{3}}-\eta_{a_{2}}\bar{\eta}_{a_{1}}D_{a_{1}}\\
&&(\partial_{a_{3}}\nabla_{12}) \rightarrow (\partial_{a_{3}}\nabla_{12})+\eta_{a_{3}}\bar{\eta}_{a_{1}}D_{a_{1}}-\eta_{a_{3}}\bar{\eta}_{a_{2}}D_{a_{2}}.
\end{eqnarray}

and can write

\begin{eqnarray}
\mathcal {A}(\Phi(z)) = \int d\eta_{a_{1}}d\bar{\eta}_{a_{1}}d\eta_{a_{2}}d\bar{\eta}_{a_{2}}d\eta_{a_{3}}d\bar{\eta}_{a_{3}}
                        (1+\eta_{a_{1}}\bar{\eta}_{a_{1}})(1+\eta_{a_{2}}\bar{\eta}_{a_{2}})(1+\eta_{a_{3}}\bar{\eta}_{a_{3}})\nonumber\\
                        exp \hat W \Phi(z_1;a_{1})\Phi(z_2;a_{2})\Phi(z_3;a_{3})\mid_{a_{1}=a_{2}=a_{3}=0}\label{exp1}
\end{eqnarray}

where

\begin{eqnarray}
\hat W = \frac{1}{2}[1+\lambda^{2}(\partial_{a_{1}}\partial_{a_{2}}+\frac{1}{4}\eta_{a_{1}}\bar{\eta}_{a_{2}}\Box_{a_{2}}+\frac{1}{4}\eta_{a_{2}}\bar{\eta}_{a_{1}}\Box_{a_{1}})]
[\partial_{a_{3}}\nabla_{12}+\eta_{a_{3}}\bar{\eta}_{a_{1}}D_{a_{1}}-\eta_{a_{3}}\bar{\eta}_{a_{2}}D_{a_{2}}]\nonumber\\
+\frac{1}{2}[1+\lambda^{2}(\partial_{a_{2}}\partial_{a_{3}}+\frac{1}{4}\eta_{a_{2}}\bar{\eta}_{a_{3}}\Box_{a_{3}}+\frac{1}{4}\eta_{a_{3}}\bar{\eta}_{a_{2}}\Box_{a_{2}})]
[\partial_{a_{1}}\nabla_{23}+\eta_{a_{1}}\bar{\eta}_{a_{2}}D_{a_{2}}-\eta_{a_{1}}\bar{\eta}_{a_{3}}D_{a_{3}}]\nonumber\\
+\frac{1}{2}[1+\lambda^{2}(\partial_{a_{3}}\partial_{a_{1}}+\frac{1}{4}\eta_{a_{3}}\bar{\eta}_{a_{1}}\Box_{a_{1}}+\frac{1}{4}\eta_{a_{1}}\bar{\eta}_{a_{3}}\Box_{a_{3}})]
[\partial_{a_{2}}\nabla_{31}+\eta_{a_{2}}\bar{\eta}_{a_{3}}D_{a_{3}}-\eta_{a_{2}}\bar{\eta}_{a_{1}}D_{a_{1}}]\label{SagnottiW}
\end{eqnarray}
This operator generates all terms in the cubic interaction of any three HS fields with any possible number of derivatives $\Delta$ in the range $\Delta_{min}\leq \Delta \leq \Delta_{max}$. Another possible form of the $\hat W$ operator is

\begin{eqnarray}
\hat W = [1+\lambda^{2}(\partial_{a_{1}}\partial_{a_{2}}+\frac{1}{2}\eta_{a_{1}}\bar{\eta}_{a_{2}}\Box_{a_{2}})]
[(\partial_{a_{3}}\nabla_{1})+\frac{1}{2}\eta_{a_{3}}\bar{\eta}_{a_{1}}D_{a_{1}}-\frac{1}{2}\eta_{a_{3}}\bar{\eta}_{a_{2}}D_{a_{2}}
+\frac{1}{2}\eta_{a_{3}}\bar{\eta}_{a_{3}}D_{a_{3}}]\nonumber\\
+[1+\lambda^{2}(\partial_{a_{2}}\partial_{a_{3}}+\frac{1}{2}\eta_{a_{2}}\bar{\eta}_{a_{3}}\Box_{a_{3}})]
[(\partial_{a_{1}}\nabla_{2})+\frac{1}{2}\eta_{a_{1}}\bar{\eta}_{a_{2}}D_{a_{2}}-\frac{1}{2}\eta_{a_{1}}\bar{\eta}_{a_{3}}D_{a_{3}}
+\frac{1}{2}\eta_{a_{1}}\bar{\eta}_{a_{1}}D_{a_{1}}]\nonumber\\
+[1+\lambda^{2}(\partial_{a_{3}}\partial_{a_{1}}+\frac{1}{2}\eta_{a_{3}}\bar{\eta}_{a_{1}}\Box_{a_{1}})]
[(\partial_{a_{2}}\nabla_{3})+\frac{1}{2}\eta_{a_{2}}\bar{\eta}_{a_{3}}D_{a_{3}}-\frac{1}{2}\eta_{a_{2}}\bar{\eta}_{a_{1}}D_{a_{1}}
+\frac{1}{2}\eta_{a_{2}}\bar{\eta}_{a_{2}}D_{a_{2}}]\label{MMRW}
\end{eqnarray}
This case generates the Lagrangian derived in \cite{GTI}. The leading term of that Lagrangian is (\ref{leading}). These two operators (\ref{SagnottiW}) and (\ref{MMRW}) generate two Lagrangians that differ from each other just by partial integration and field redefinition. All interactions of HS gauge fields with any number of derivatives are unique and are generated by both operators (\ref{SagnottiW}) and (\ref{MMRW}).

In the case of (\ref{SagnottiW}) we have
\begin{eqnarray}
\setlength{\unitlength}{0.254mm}
\begin{picture}(280,275)(125,-355)
        \allinethickness{0.254mm}\path(405,-80)(405,-175) 
        \allinethickness{0.254mm}\path(345,-235)(345,-80) 
        \allinethickness{0.254mm}\path(285,-295)(285,-80) 
        \allinethickness{0.254mm}\path(225,-355)(225,-80) 
        \allinethickness{0.254mm}\path(125,-80)(125,-355) 
        \allinethickness{0.254mm}\path(165,-80)(165,-355) 
        \allinethickness{0.254mm}\path(405,-80)(125,-80) 
        \allinethickness{0.254mm}\path(405,-175)(125,-175) 
        \allinethickness{0.254mm}\path(345,-235)(125,-235) 
        \allinethickness{0.254mm}\path(285,-295)(125,-295) 
        \allinethickness{0.254mm}\path(225,-355)(125,-355) 
        \allinethickness{0.254mm}\path(125,-115)(405,-115) 
        \allinethickness{0.254mm}\path(125,-80)(165,-115) 
        \put(130,-110){\shortstack{$\Box_{a_{i}}$}} 
        \put(141,-94){\shortstack{$D_{a_{i}}$}} 
        \put(190,-106){\shortstack{$0$}} 
        \put(250,-106){\shortstack{$1$}} 
        \put(310,-106){\shortstack{$2$}} 
        \put(370,-106){\shortstack{$3$}} 
        \put(140,-151){\shortstack{$0$}} 
        \put(140,-211){\shortstack{$1$}} 
        \put(140,-271){\shortstack{$2$}} 
        \put(140,-331){\shortstack{$3$}} 
        \put(185,-151){\shortstack{$\mathcal {A}^{00}$}} 
        \put(250,-151){\shortstack{$0$}} 
        \put(295,-151){\shortstack{$\mathcal {A}^{20}$}} 
        \put(370,-151){\shortstack{$0$}} 
        \put(190,-211){\shortstack{$0$}} 
        \put(185,-271){\shortstack{$\mathcal {A}^{02}$}} 
        \put(185,-331){\shortstack{$\mathcal {A}^{03}$}} 
        \put(235,-211){\shortstack{$\mathcal {A}^{11}$}} 
        \put(295,-211){\shortstack{$\mathcal {A}^{21}$}} 
        \put(250,-271){\shortstack{$\mathcal {A}^{12}$}} 
\end{picture}
\end{eqnarray}
In the case of (\ref{MMRW}) we have
\begin{eqnarray}
\setlength{\unitlength}{0.254mm}
\begin{picture}(280,275)(125,-355)
        \allinethickness{0.254mm}\path(405,-80)(405,-175) 
        \allinethickness{0.254mm}\path(345,-235)(345,-80) 
        \allinethickness{0.254mm}\path(285,-295)(285,-80) 
        \allinethickness{0.254mm}\path(225,-355)(225,-80) 
        \allinethickness{0.254mm}\path(125,-80)(125,-355) 
        \allinethickness{0.254mm}\path(165,-80)(165,-355) 
        \allinethickness{0.254mm}\path(405,-80)(125,-80) 
        \allinethickness{0.254mm}\path(405,-175)(125,-175) 
        \allinethickness{0.254mm}\path(345,-235)(125,-235) 
        \allinethickness{0.254mm}\path(285,-295)(125,-295) 
        \allinethickness{0.254mm}\path(225,-355)(125,-355) 
        \allinethickness{0.254mm}\path(125,-115)(405,-115) 
        \allinethickness{0.254mm}\path(125,-80)(165,-115) 
        \put(130,-110){\shortstack{$\Box_{a_{i}}$}} 
        \put(141,-94){\shortstack{$D_{a_{i}}$}} 
        \put(190,-106){\shortstack{$0$}} 
        \put(250,-106){\shortstack{$1$}} 
        \put(310,-106){\shortstack{$2$}} 
        \put(370,-106){\shortstack{$3$}} 
        \put(140,-151){\shortstack{$0$}} 
        \put(140,-211){\shortstack{$1$}} 
        \put(140,-271){\shortstack{$2$}} 
        \put(140,-331){\shortstack{$3$}} 
        \put(185,-151){\shortstack{$\mathcal {A}^{00}$}} 
        \put(235,-151){\shortstack{$\mathcal {A}^{10}$}} 
        \put(295,-151){\shortstack{$\mathcal {A}^{20}$}} 
        \put(355,-151){\shortstack{$\mathcal {A}^{30}$}} 
        \put(190,-211){\shortstack{$0$}} 
        \put(190,-271){\shortstack{$0$}} 
        \put(185,-331){\shortstack{$\mathcal {A}^{03}$}} 
        \put(235,-211){\shortstack{$\mathcal {A}^{11}$}} 
        \put(295,-211){\shortstack{$\mathcal {A}^{21}$}} 
        \put(235,-271){\shortstack{$\mathcal {A}^{12}$}} 
\end{picture}
\end{eqnarray}
Both forms of the same cubic Lagrangian are very useful for further investigations.
\subsection*{Acknowledgements}
The authors are grateful to A. Sagnotti and R. Mkrtchyan for discussions on the subject of this work.
This work is supported in part by Alexander von Humboldt Foundation under 3.4-Fokoop-ARM/1059429.
Work of K.M. was made with partial support of CRDF-NFSAT-SCS MES RA ECSP 09\_01/A-31.
\\

\noindent {\bf Note added.} When the present work was on its final
stage for submission, the paper \cite{Fotopoulos:2010ay} appeared in the archive which includes an
analysis of off-shell cubic interactions in HS field theories from String theory point of view, using BRST technique.
The two results are closely related.


\begin{thebibliography}{100}
\bibitem{GTI}
  R.~Manvelyan, K.~Mkrtchyan and W.~R\"uhl,
    "General trilinear interaction for arbitrary even higher spin gauge fields",
      Nucl.\ Phys.\  B {\bf 836} (2010) 204, arXiv:1003.2877 [hep-th].
\bibitem{ST}
A. Sagnotti, M. Taronna, ``String Lessons for Higher-Spin Interactions.''
arXiv:1006.5242 [hep-th]
\bibitem{VasilievEqn}
  M.~A.~Vasiliev, ``Consistent equation for interacting gauge fields of all spins in (3+1)-dimensions.'',
  Phys. Lett. B {\bf 243} (1990) 378-382. M.~A.~Vasiliev, `` Nonlinear equations for symmetric massless higher spin fields in $(A)dS_{d}$.''
  Phys. Lett. B {\bf 567} (2003) 139-151, arXiv:hep-th/0304049.
\bibitem{Frons}
    C.~Fronsdal, ``Singletons And Massless, Integral Spin Fields On De Sitter Space (Elementary
    Particles In A Curved Space Vii),'' Phys.\ Rev.\ D {\bf 20},
    (1979) 848;``Massless Fields With Integer Spin,'' Phys.\ Rev.\ D {\bf 18} (1978) 3624.
\bibitem{MMR0}
  R.~Manvelyan, K.~Mkrtchyan and W.~R\"uhl,
  ``Direct construction of a cubic selfinteraction for higher spin gauge
  fields,'' Nucl. Phys. B {\bf 844}(2011) 348, [arXiv:1002.1358 [hep-th]].
\bibitem{M}
  R.~Manvelyan, K.~Mkrtchyan and W.~R\"uhl,
  ``Off-shell construction of some trilinear higher spin gauge field
  interactions,''
  Nucl.\ Phys.\  B {\bf 826} (2010) 1, [arXiv:0903.0243 [hep-th]].
\bibitem{MM}
R.~Manvelyan and K.~Mkrtchyan,
``Conformal invariant interaction of a scalar field with the higher spin
  field in $AdS_{D}$,'' Mod. Phys. Lett. A {\bf 25} (2010)1333, [arXiv:0903.0058 [hep-th]].
\bibitem{MR}
R.~Manvelyan and W.~R\"uhl, ``Conformal coupling of higher spin
gauge fields to a scalar field in AdS(4) and generalized Weyl
invariance,'' Phys.\ Lett.\ B {\bf 593} (2004) 253,
[arXiv:hep-th/0403241].
\bibitem{Vasiliev}
  E.~S.~Fradkin and M.~A.~Vasiliev, ``On The Gravitational Interaction
  Of Massless Higher Spin Fields,'' Phys.\ Lett.\ B {\bf 189} (1987)
  89.
\bibitem{boulanger}
  Nicolas Boulanger, Serge Leclercq, Per Sundell,
  ``On The Uniqueness of Minimal Coupling in Higher-Spin Gauge Theory,''
  JHEP 0808:056,2008;  [arXiv:0805.2764 [hep-th]].
\bibitem{vanDam}
  F.~A.~Berends, G.~J.~H.~Burgers and H.~van Dam,
  ``Explicit Construction Of Conserved Currents For Massless Fields Of
  Arbitrary Spin,'' Nucl.\ Phys.\  B {\bf 271} (1986) 429;
   F.~A.~Berends, G.~J.~H.~Burgers and H.~Van Dam,
  ``On Spin Three Selfinteractions,''
  Z.\ Phys.\  C {\bf 24} (1984) 247;
  F.~A.~Berends, G.~J.~H.~Burgers and H.~van Dam,
  ``On The Theoretical Problems In Constructing Interactions Involving Higher
  Spin Massless Particles,''
  Nucl.\ Phys.\  B {\bf 260} (1985) 295.
\bibitem{bek}
Xavier Bekaert, Nicolas Boulanger and Serge Leclercq,
``Strong obstruction of the Berends-Burgers-van Dam spin-3 vertex.''
J.Phys.A43:185401,2010. arXiv:1002.0289 [hep-th].
\bibitem{Metsaev}
  R.~R.~Metsaev,
  ``Cubic interaction vertices for massive and massless higher spin fields,''
  Nucl.\ Phys.\  B {\bf 759} (2006) 147
  [arXiv:hep-th/0512342];
  R.~R.~Metsaev,  ``Cubic interaction vertices for fermionic and bosonic arbitrary spin  fields,''
  arXiv:0712.3526 [hep-th].
\bibitem{Vasiliev1}
  E.~S.~Fradkin and M.~A.~Vasiliev, ``Cubic Interaction In
  Extended Theories Of Massless Higher Spin Fields,'' Nucl.\ Phys.\ B
  {\bf 291} (1987) 141.
  M.~A.~Vasiliev, "Cubic Interactions of Bosonic Higher Spin Gauge Fields in $AdS_{5}$",
  [arXiv:hep-th/0106200].
  K. B. Alkalaev, M.~A.~Vasiliev, "N = 1 Supersymmetric Theory of Higher Spin Gauge Fields in $AdS_{5}$ at the Cubic Level",
  [arXiv:hep-th/0206068]
\bibitem{Damour}
 T.~Damour and S.~Deser,
  ``Higher derivative interactions of higher spin gauge fields,''
  Class.\ Quant.\ Grav.\  {\bf 4}, L95 (1987).
\bibitem{Petkou}
  A.~Fotopoulos, N.~Irges, A.~C.~Petkou and M.~Tsulaia,
  ``Higher-Spin Gauge Fields Interacting with Scalars: The Lagrangian Cubic
  Vertex,''
  JHEP {\bf 0710} (2007) 021;
  [arXiv:0708.1399 [hep-th]].
  I.~L.~Buchbinder, A.~Fotopoulos, A.~C.~Petkou and M.~Tsulaia,
  ``Constructing the cubic interaction vertex of higher spin gauge fields,''
  Phys.\ Rev.\  D {\bf 74} (2006) 105018;
  [arXiv:hep-th/0609082].
    A. Fotopoulos and M. Tsulaia,
    ``Current Exchanges for Reducible Higher Spin Modes on AdS.'' arXiv:1007.0747 [hep-th]
\bibitem{polyakov}
Dimitri Polyakov, ``Gravitational Couplings of Higher Spins from String Theory.''
arXiv:1005.5512 [hep-th],
 ``Interactions of Massless Higher Spin Fields From String Theory.''
arXiv:0910.5338 [hep-th].
\bibitem{Zinoviev}
Yu. M. Zinoviev, ``Spin 3 cubic vertices in a frame-like formalism.''
JHEP 1008:084,2010; arXiv:1007.0158 [hep-th]
\bibitem{review}
    M. A. Vasiliev, ``Higher Spin Gauge Theories in Various Dimensions'',
    Fortsch. Phys. 52, 702 (2004) [arXiv:hep-th/0401177].
    X. Bekaert, S. Cnockaert, C. Iazeolla and M. A. Vasiliev,
    ``Nonlinear higher spin theories in various dimensions'', [arXiv:hep-th/0503128].
    D. Sorokin,``Introduction to the Classical Theory of Higher Spins'' AIP Conf. Proc. 767, 172
    (2005); [arXiv:hep-th/0405069].
    N. Bouatta, G. Compere and A. Sagnotti, ``An Introduction to Free Higher-Spin Fields''; [arXiv:hep-th/0409068].
    Xavier Bekaert, Nicolas Boulanger and Per Sundell,
    ``How higher-spin gravity surpasses the spin two barrier: no-go theorems versus yes-go examples.'' arXiv:1007.0435 [hep-th]
\bibitem{Savvidy}
  G.~Savvidy,
  ``Connection between non-Abelian tensor gauge fields and open strings,''
  J.\ Phys.\ A  {\bf 42} (2009) 065403
  [arXiv:0808.2244 [hep-th]];
  I.~Antoniadis and G.~Savvidy,
  ``Scattering of charged tensor bosons in gauge and superstring theories,''
  Fortsch.\ Phys.\  {\bf 58} (2010) 811
  [arXiv:0907.3553 [hep-th]].
\bibitem{MMR1}
R.~Manvelyan, K.~Mkrtchyan and W.~R\"uhl,
  ``Ultraviolet behaviour of higher spin gauge field propagators and one loop
  mass renormalization,''
  Nucl.\ Phys.\  B {\bf 803} (2008) 405
  [arXiv:0804.1211 [hep-th]].
\bibitem{Fotopoulos:2010ay}
  A.~Fotopoulos and M.~Tsulaia,
  ``On the Tensionless Limit of String theory, Off - Shell Higher Spin
  Interaction Vertices and BCFW Recursion Relations,''
  arXiv:1009.0727 [hep-th].


\end{thebibliography}
\end{document}